\documentclass[letterpaper, 10 pt, conference]{ieeeconf}  

\IEEEoverridecommandlockouts                              
\usepackage{url}

\usepackage{pdflscape}
\usepackage{afterpage}
\usepackage[T1]{fontenc}

\usepackage[english]{babel}
\usepackage[utf8x]{inputenc}
\usepackage{authblk}
\usepackage{graphics} 
\usepackage{amsmath}
\usepackage{url}
\usepackage{graphicx}

\usepackage{afterpage}

\usepackage{longtable} 
\usepackage{pdflscape} 

\usepackage{listings}

\title{\LARGE \bf
Event Detection on Twitter
}

\author{Ozlem Ceren Sahin$^{1}$, Nesime Tatbul$^{2}$, Pinar Karagoz$^{1}$

\thanks{$^{1}$O. C. Sahin is with Computer Engineering Department,
        Middle East Technical University, Turkey and Huawei Technologies Co. Ltd.
        {\tt\small e1746668 at ceng.metu.edu.tr}}%
        \thanks{$^{2}$P. Karagoz is with Computer Engineering Department,
        Middle East Technical University, Turkey
        {\tt\small karagoz at ceng.metu.edu.tr}}%
\thanks{$^{3}$Nesime Tatbul is with CSAIL,
MIT, USA
        {\tt\small tatbul at csail.mit.edu.tr}}%
}

\begin{document}

\maketitle
\thispagestyle{empty}
\pagestyle{empty}

\begin{abstract}

Detecting events by using social media has been an active research problem. In this work, we investigate and compare the performance of two methods for event detection in Twitter by using Apache Storm as the stream processing infrastructure. The first event detection method is based on identifying uncommonly common words inside tweet blocks, and the second one is based on clustering tweets to detect a cluster as an event. Each of the methods has its own characteristics. Uncommonly common word based method relies on the burst of words and hence is not affected from concurrency problems in distributed environment. On the other hand, clustering based method includes a finer grained analysis, but it is sensitive to the concurrent processing. We investigate the effect of stream processing and concurrency handling support provided by Apace Storm on event detection by these methods.  

\textit{Keywords- Event Detection, Micro-blogging, Twitter, Semantics, Real-time Evaluation, Distributed System, Uncommonly Common Algorithm, Hierarchical Clustering}
\end{abstract}


\section{INTRODUCTION}
Due to the performance constraints we use Apache Storm as real-time distributed computation system which is originally created by Nathan Marz \cite{c1} and team at BackType \cite{c2} and become open source after being acquired by Twitter \cite{c3}. Storm makes processing unbounded streams of data in real-time possible and it can be used in any programming language. Some of the use cases of Storm are real-time analytics, online machine learning and distributed remote procedure call. The main advantage of Storm is its efficiency, i.e Storm can process a million tuples per second per node as a benchmark. Besides Storm is scalable, fault-tolerant, guarantees data will be processed and is easy to set up and operate.

\section{RELATED WORK}

Event detection, also known as event tracking, detects events by processing textual materials like social media \cite{c4}. After Twitter comes into scene in 2006 it is used by millions of users around the world and many studies use Twitter data for information retrieval \cite{c5}. An implementation of real-time event detection from Twitter is described in \cite{c6}. In that work, Sankaranarayanan et al. receive tweets of specified users from different parts of the world; cluster them for event detection, and assign a geographic location to the event in order to display it on a map. In a similar study, Sakaki detects earthquakes and their locations by following tweets \cite{c7}. There are also some improvements about event detection on Twitter, for example returning the first tweet posted about an event \cite{c8}. In another work, Park et al. detect events related to a baseball game and lists people who watch the game on TV \cite{c9}. In several studies semantics in word co-occurrences has been used to find the similarities. In the study "Dynamic Relationship and Event Discovery" \cite{c10}, burst detection and co-occurrence methods are used for event detection. They prepare lists to identify words which belongs to same event. To achieve this they use the burst approach, i.e. if two words have burst in same window, they point the same event. A similar approach is used in "Event Detection and Tracking in Social Streams" \cite{c11}. They generate a graph where terms as represented as nodes and co-occurrences are represented as edges, then connected sub-graphs are labeled as event clusters.

Methods identifying hashtag or word associations are also presented in some studies. They aim to detect hashtags with an increasing co-occurrence with a given in a period of time. Another example of analyzing term associations \cite{c12}. In another work, online store aims to predict the name of the product sold on online-store by using previous queries \cite{c13}.

Ozdikis, Senkul and Oguztuzun proposed a method to enhance the event detection techniques by using lexico-semantic expansion of tweets. To achieve this they use document similarity techniques and clustering algorithms \cite{c14}.

\section{METHOD}

As enhancement we have proposed three methods:
\begin{itemize}
\item \textbf{Preprocess data with NLP} to increase accuracy. 
\item \textbf{Apache Storm} for real-time twitter computation. 
\item Event detection with regard to \textbf{geolocation}. 
\end{itemize}

\subsection{Data Collection}
Twitter provides REST APIs \cite{c15} and Streaming APIs \cite{c16} so that users can get Twitter data through these channels. REST API is used to send request to have information about tweets, users, locations or other objects of Twitter data and get JSON or XML formatted responses as common. On the other hand Streaming API provides a stream of Twitter data which can be filtered by a desired criteria. 

During our project we have used Twitter Streaming API, named Twitter4j \cite{c17} which is a Java library, and get data from USA and Canada by using location filter which works with longitude and latitude values. So that Twitter sends some of the USA and Canada data through the stream created by Twitter4j API. By using this API, we have collected millions of tweets from many users and about many topics which are processed in our experiments to detect the events. 

 In our experiments, we used nearly 12M  tweets  gathered for one week from May 31, 2016 to April 7, 2016. These tweets are preprocessed during collection and saved to Apache Cassandra database with document numbers. Document numbers are needed for uncommonly common algorithm since this methodology identifies the bursty keywords between documents. For our methodology, documents correspond to the collection of tweets for 6-minute interval. For example, document 1 includes the tweets from May 31, 2016 - 00:00 to May 31, 2016 00:06; document 2 includes the tweets from May 31, 2016 - 00:06 to May 31, 2016 00:12 and so on.
 
\subsection{Preprocessing}

Before execution of the event detection methods we need to preprocess data so that we can increase the accuracy and performance. 
\begin{itemize}
\item \textbf{Tokenize} the sentence into words using spaces and punctuation. By doing so, we can keep the word count and calculate tf-idf values for most common words.
\item Apply \textbf{stemming and stop word elimination} using NLP library \cite{c18}. The reason for stemmer and stop word elimination is increasing the accuracy and performance. Since stop words are seen in most of the tweets, they act as continuous events seen all the time, but they should not be evaluated as events.
\item \textbf{Remove} sentences started with "I am at" and URLs since they do not correspond to a event. 
\item \textbf{Remove} characters from a word which repeats more than 2 times consecutively. For example replace "noooooooooooooo" with "no". Stemming and removing repeating characters are useful in text mining because the word "firing", "fireeeee" and "fire" should be counted as same word. Otherwise it will increase the possibility of missing the event.
\end{itemize}

\subsection{Introduction To Apache Storm}
For an efficient and reliable computation, we have used Apache Storm in our project. Apache Storm can be used in local mode during implementation since it creates Storm nodes on local machine and makes implementation and debugging easier or it can be used in cluster mode which enables the distributed computation. When Storm is used in cluster mode, it creates distributed clusters and manage them automatically after their configurations are done.

There are just three abstractions in Storm: spouts, bolts, and topologies. 
\begin{itemize}
\item 
Storm uses "spouts" as source of streams in a computation. Spout even reads from a queuing broker such as Kafka or RabbitMQ or generates its own stream or read from somewhere like the Twitter streaming API or from a database like Apache Cassandra. For our project we replicate the tweets stored in Apache Cassandra database in time order.
\item
Storm uses "bolt" to process any number of input streams and produce any number of new output streams. Most of the logic of a computation goes into bolts, such as functions, filters, streaming joins, streaming aggregations, talking to databases, and so on.
\item
Storm uses "topology" as a network of spouts and bolts, with each edge in the network representing a bolt subscribing to the output stream of some other spout or bolt. A topology is an arbitrarily complex multi-stage stream computation. Topologies run indefinitely when deployed. 
\newline
Each spout or bolt executes as many tasks across the cluster. Each task corresponds to one thread of execution, and stream groupings define how to send tuples from one set of tasks to another set of tasks. Parallelism for each spout or bolt is set by developer.
\newline

\begin{figure} [ht!]
  \includegraphics[width=8cm,height=5cm]{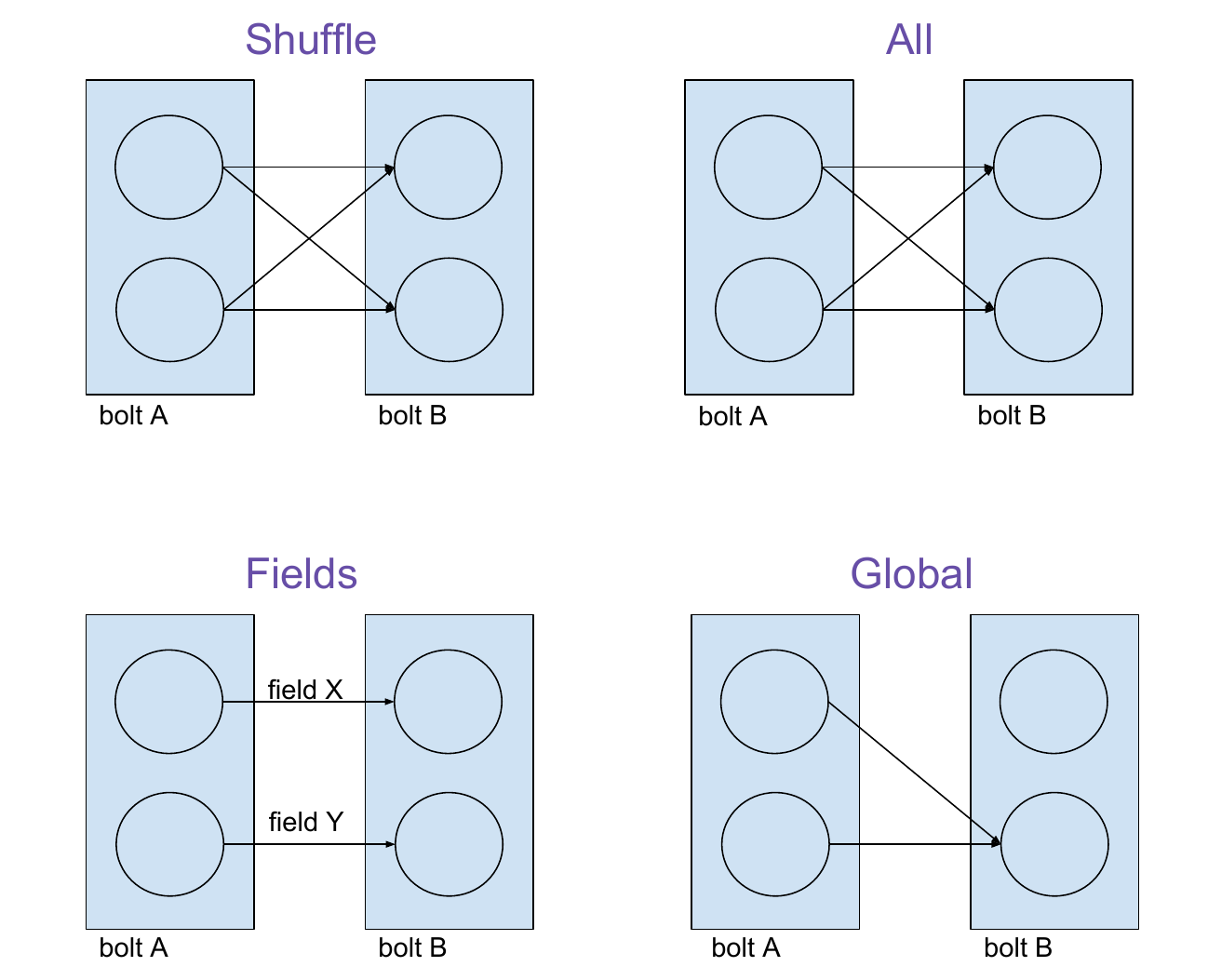}
  \caption{Storm grouping types (taken from \cite{c22})}
\end{figure}

Apache storm provides 5 grouping types, 4 of them can be seen in Figure 1:
\begin{itemize}
\item Shuffle grouping: This type of grouping distributes stream among bolt tasks randomly and fairly. Distribution is handled by Apache Storm.
\item All grouping: This type of grouping replicates stream to all bolt tasks. 
\item Fields grouping: This type of grouping partitions stream on a user-specified field. 
\item Global grouping: This type of grouping gets entire stream to single task. 
\item Direct grouping: This is a special kind of grouping. A stream grouped this way means that the producer of the tuple decides which task of the consumer will receive this tuple. Therefore distribution of tuples fairly among tasks depends on the developer.
\end{itemize}
\end{itemize}

The terms described above will be used in the rest of this paper.

\subsection{Uncommonly common algorithm for event detection}

\begin{figure} [ht!]
  \includegraphics[width=9cm,height=3cm]{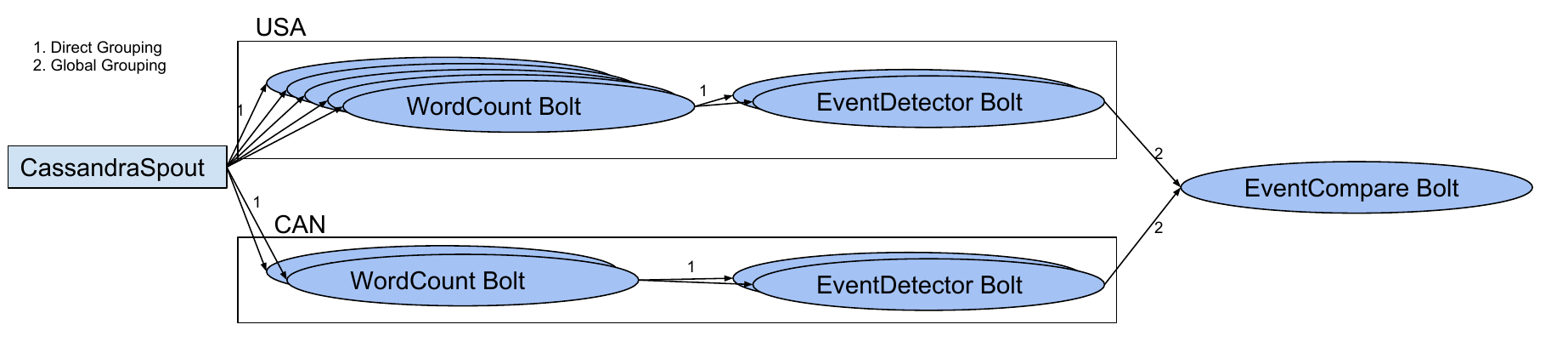}
  \caption{Storm topology with Direct Grouping}
\end{figure}

\begin{figure} [ht!]
  \includegraphics[width=9cm,height=3cm]{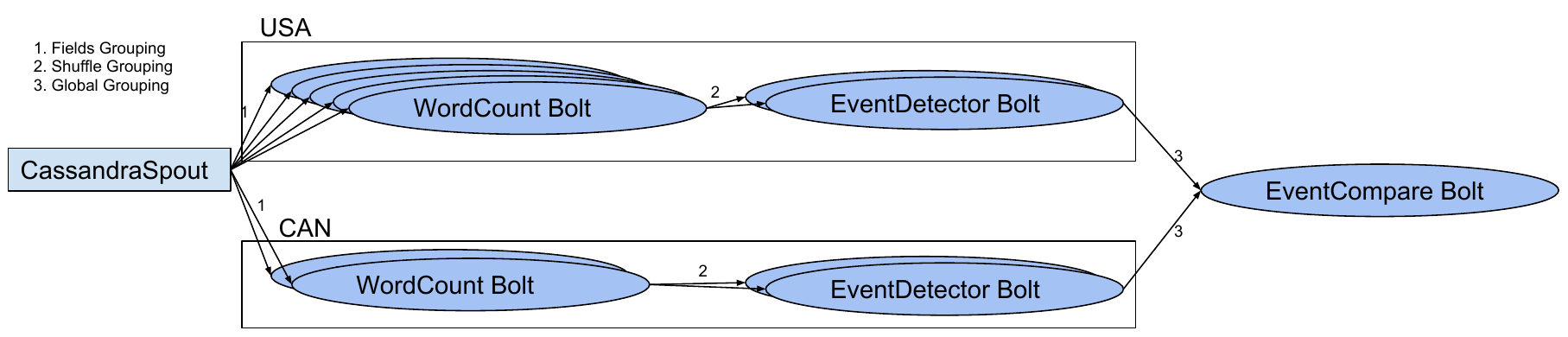}
  \caption{Storm topology with Sleep}
\end{figure}

\subsubsection{User defined parameters}
There are 2 parameters defined by the user:
\begin{itemize}
\item tf-idf\_event\_rate: This parameter is used to define the increment rate of tf-idf value between the last two documents to be an event.
\item common\_word\_threshold: This parameter defines the threshold number which is used to presume a word as common word. Only the common words are subject to the uncommonly common algorithm.
\end{itemize}

\subsubsection{Source Of Stream}
As a source of stream, our methodology replicates the data saved to Apache Cassandra database in document order. When replication of tweets inside a document is finished, spout does not start replication of the next document immediately. Next document replication is suspended until the process of the current document is completed. The reason of suspension between document processes is the reliability of the system and the correctness of the event candidates on current document. If the next document streaming is started before the current document has finished processing, there is a high possibility that two documents are interfered which will result in wrong event candidates for current document. Therefore, it is needed to define a streaming protocol which handles suspension. Apache Storm does not provide a built-in methodology for this purpose, hence in our project we describe two different approaches. First one is sleeping spout for some time determined experimentally and the second one is using direct grouping defined by Storm. 

Our first approach uses shuffle grouping and fields grouping where scheduling and tuple distribution is handled by Apache Storm. This approach leads to a more efficient distribution and processing time than other approach. However as a disadvantage of this approach, it is not possible to figure out whether the process of current document is ended or not. Therefore spout needs to wait(sleep) for some time to guarantee that the documents are not interfering.

Second approach uses direct grouping and direct streaming where tuples are distributed in turn to each task manually by developer. Hereby distribution and scheduling is not well-handled as first approach, because the availability of the tasks can not be controlled during distribution. On the other hand the advantage of this method is not losing time by sleeps. 

\subsubsection{Step 1 - Count Words}
Streaming tuples firstly splitted into words and emitted to word count bolt by fields grouping. Tasks of word count bolt simply count the number of words for the current document and eliminates rare words which has count less than common\_word\_threshold for performance. By this pre-elimination we can define the most common words in current document and avoid unnecessary calculations for the uncommon words. 

\subsubsection{Step 2 - Calculating TF-IDF}
Tf-idf is the short form of term frequency-inverse document frequency which shows a numerical statistic pointing out the importance of a word to a document in a collection of documents. This technique is mostly used in information retrieval and text mining as a weighting factor \cite{c10}.

The value tf in a document and idf among all documents can be calculated as follows:
\begin{equation}
  \mathrm{tf}(t, d) =\cfrac{f_{t,d}}{|\{ t' \in d \}|}
\end{equation}
where,
\begin{itemize}
\item $f_{t,d}$: the number of times that term t occurs in document d.
\item $|\{ t' \in d \}|$: total number of terms in document d.
\end{itemize}

\begin{equation}
  \mathrm{idf}(t, D) = log \cfrac{N}{1 + |\{ d \in D : t \in d \}|}
\end{equation}
where,
\begin{itemize}
\item N: total number of documents in the corpus N = \{|D|\}
\item |\{${d\in D:t\in d}$\}|: number of documents where the term t appears. Addition of 1 prevents division-by-zero.

\end{itemize}

And finally the over all tf-idf value can be computed as follows:
\begin{equation}
    \mathrm{tf-idf}(t,d,D) = \mathrm{tf}(t,d) \cdot \mathrm{idf}(t, D) 
\end{equation}
where,
\begin{itemize}
\item $\mathrm{tf}(t,d)$   : tf value calculated by using equation 1
\item $\mathrm{idf}(t, D)$ : idf  value calculated by using equation 2
\end{itemize}
\  \\
    
Our methodology computes the tf-idf value for each common word for the current and previous documents and calculates the increment rate between last two documents instead of using the tf-idf value of only the current document.

\begin{equation}
  \mathrm{increment\_rate} =\cfrac{ tfidf\_current}{tfidf\_previous}
\end{equation}
where,
\begin{itemize}
\item $ tf-idf\_current$: the tfidf value of the word for current document.
\item $ tf-idf\_previous$: the tfidf value of the word for previous document.
\end{itemize}

To sign a word as an event candidate, the increment\_rate of the word should be above the specified threshold which is defined by tf-idf\_event\_rate. To sign a word as uncommonly-common word (also event candidate), this word should have a high tf-idf value only for the last document. If the tf-idf value of a word is high or low for each document, it means there is no current burst on that word which means it does not show an event for the current document.

\subsubsection{Apache Storm Topology}
The equivalents of Apache Storm abstractions in our system are as follows:
\begin{itemize}
\item 
Spout: The name of the spout in this approach is CassandraSpout which reads tweets from Apache Cassandra database one by one in order, split tweets into words and emits them to next bolt. 
\item Count Bolts: These bolts are responsible to hold count of each word separately. This bolt emits the word to next bolt for tfidf calculation phase once when count of the word reached to the threshold immediately.  
\item Event Detector Bolt: This bolt stores the words emitted from word count bolt and calculates the tf-idf value of the last two documents and increment rates for each common word at the end of the document. 
However, the methodology using sleep is different. Since it does not aware of the end of the document, it immediately calculates the tf-idf values of common words emitted to this bolt. Therefore this methodology is not appropriate for the real-time systems. 
\item Event Compare Bolt: This bolt stores event candidate words in Apache Cassandra database and creates line charts for the event candidate word by using the tf-idf value of the word for the last 10 document. 
\item The Apache Storm topology of Twitter Event Detection can be seen in Figure 2 and Figure 3 with direct grouping and with sleep respectively.
\newline

\end{itemize}

\subsection{Hierarchical clustering algorithm for event detection}

\begin{figure} [ht!]
  \includegraphics[width=9cm,height=3cm]{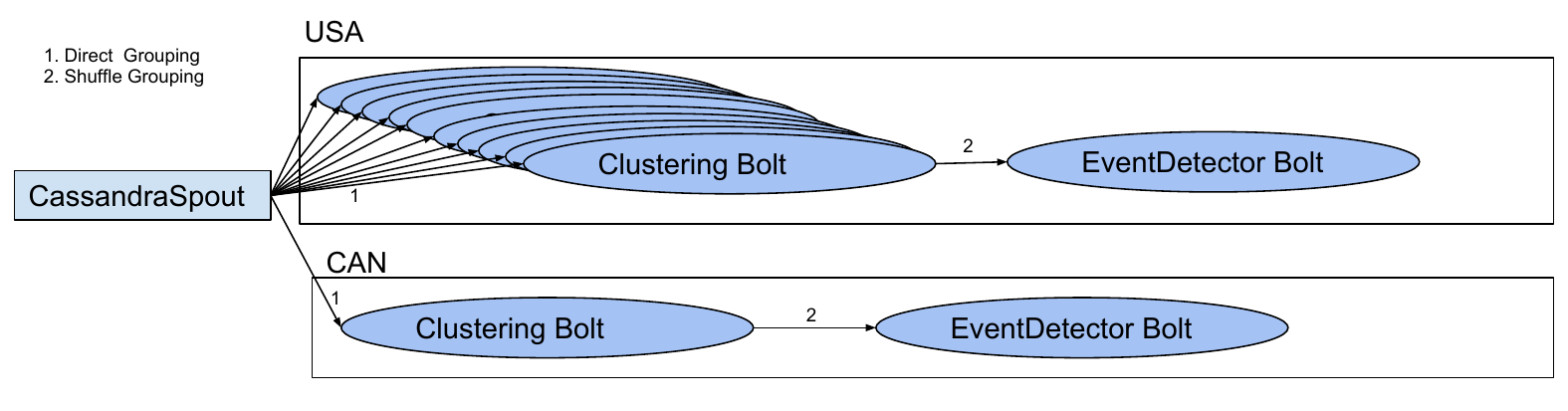}
  \caption{Clustering topology}
\end{figure}

\subsubsection{User defined parameters}
There are 1 parameter defined by the user:
\begin{itemize}
\item num\_tweet\_threshold: During the computation there are many clusters created. Therefore it is not efficient to evaluate each cluster at the end of each document and there should be a threshold which points out the minimum number of tweets required in a cluster. 
\end{itemize}

\subsubsection{Source Of Stream}
CassandraSpout of this method gets tweets from Apache Cassandra database in tweet-time order and emit them by direct grouping to the next bolt similar to the previous method. Between document replications spout waits until all tasks finish their process for current document. Spout of this approach directly emits the vectorized tweet since clustering is based on the cosine similarity of the whole tweet vector. Tweet vector shows the words inside a tweet with the normalized weight as a map. For example the vector of the tweet "RIP Muhammed Ali RIP" is \{"RIP":0.5, "Muhammed":0.25, "Ali":0.25\}. Clusters also have the same map representations as tweets to ease the calculation of cosine similarity between tweets and clusters. 
 
\subsubsection{Hierarchical Clustering}
This methodology uses clustering algorithms in an unusual way. When the event detection procedure starts, there does not occur any clusters. During the stream of tweet vectors, new clusters are created and existing clusters are formed. At the beginning of the clustering methodology, the number of clusters is unknown (unlike to k-means clustering) and clusters are shaped during tweet streaming for each tweet which causes some problems defined below.

The related informations about formed clusters are held in Apache Cassandra tables and during the execution of this methodology, new clusters may be created, existing clusters may be updated, existing clusters may merge and some of the existing clusters may be deleted. First three operations are used to follow the growth rate of clusters where the growth rate will be used to decide even if a cluster correspond to an event or not. Deletion operation is needed for performance issues. This clustering methodology needs to get the existing clusters from database for the processing of each tweet and calculate the similarity between tweet and each cluster to update an existing one or to form a new cluster. This means for each tweet it is needed to make transaction to a fastly-grown database. Therefore it is required to remove clusters which are not active for the last 2 rounds.

As mentioned above there are four operations used on clusters: cluster creation, cluster update, cluster merge and cluster removal. Only merge and removal of clusters are related to the performance of the system. These four operations does not meet the needs related to performance and there is exponential increase of execution time as the number of clusters increase. Therefore another algorithm is evolved from the clustering method defined above and we divide clustering step into two steps. First one is sub-clustering (local clustering) inside a document which creates and updates clusters from ground base (as zero clusters occurred) until the tweets of the current document finishes. It means that when the new 6-minute document execution starts, each task of clustering bolt has zero local clusters and creates/updates local clusters during the document execution. At the end of the current document execution each task sends local clusters to the event detection bolt which merges local clusters in itself and also with global clusters from Apache Cassandra database. This methodology provides high performance improvements since it creates one database connection per document instead of each tweet(1680 vs 12M).

\subsubsection{Cosine Similarity Of Tweets}

Cosine similarity is used for two non-zero vectors to measure the angle between them. Cosine similarity of 1 means that two of the vectors are same and 0 means they are totally different. Cosine similarity is popular in a wide area of researches since it is very efficient for sparse vectors.

The cosine similarity of two vectors can be calculated as follows:
\begin{equation}
    \mathrm{cos}(\theta) = \cfrac {\sum_{i=1}^{n}{A_{i}B_{i}} } { \sqrt{\sum_{i=1}^{n}{ A_{i}^2 }} \sqrt{\sum_{i=1}^{n}{ B_{i}^2 }} }
\end{equation}
where,
\begin{itemize}
\item $A_{i}$: First tweet as vector.
\item $B_{i}$: Second tweet as vector.
\item ${cos}(\theta)$: Cosine similarity of two vectors.
\end{itemize}

This methodology computes the cosine similarity for two different operations. First one is to assign a tweet to a cluster. For that purpose cosine similarity is calculated between the current tweet vector with each local cluster vector until it finds the similar cluster. Otherwise it may create a new cluster. Second one is to merge clusters. For this merge operation cosine similarity is calculated between two local or global cluster vectors. The cosine similarity is used with a similarity threshold value while merge and update operations.

At the end of the current document execution, each cluster is evaluated to decide even if it is an event or not. For this aim the growth rate of cluster is used. The growth rate is calculated using the number of tweets which shapes a cluster using the equation below:

\begin{equation}
  \mathrm{cluster\_growth\_rate} =\cfrac{ number\_of\_tweets\_added }{total\_number\_of\_tweets}
\end{equation}
where,
\begin{itemize}
\item $ number\_of\_tweets\_added$: the number of tweets merged to cluster during the last block execution.
\item $ total\_number\_of\_tweets$: total number of tweets belongs to this cluster for all time.
\end{itemize}

To sign a cluster as an event, the cluster\_growth\_rate of the word should be greater than the specified threshold which is defined as 0.5 on our experiments.

\subsubsection{Apache Storm Topology}
In this item only the topology of two-step clustering methodology is explained:
\begin{itemize}
\item 
Spout: The name of the spout in this approach is also called as CassandraSpout which reads tweets from Apache Cassandra database one by one in time order, vectorize tweets into weighted vector maps and emits tweet vectors to next bolt. 
\item Clustering Bolts: Clustering bolts are responsible to assign tweets into local clusters by using cosine similarity. Each bolt task has zero local clusters at the beginning of the document. During the execution of current document, this bolt calculates the cosine similarity between tweet vector and local clusters to make assignment where the similarity is higher than the specified threshold or to create new cluster if cosine similarity constraint is not met for any clusters. At the end of the block, local clusters created during block streaming is emitted to the event detection bolt.
\item Event Detector Bolt: This bolt is activated at the end of the 6-min document streaming. Each clustering bolt task sends the local clusters created inside itself at the end of the block and event detector bolt gets and stores them. When all the clustering bolt sent their cluster list, this bolt starts evaluation of local clusters. This evaluation consists of two steps:

\begin{itemize}
\item Local Cluster Evaluation: First of all, this bolt collects all local clusters from each clustering bolt task and merges them in themselves. Merge operation is performed if the cosine similarity between two local clusters is bigger than or equal to the specified threshold which is 0.5. During merge operation cluster word map is updated by re-calculating the weight of each word and deleting the merged cluster. To avoid the sparse clusters, word map is examined and the words which has weight smaller than 0.01 in a cluster having more than 50 tweets are deleted from map. At the end of merge operation, all the clusters are reviewed and clusters having less than 30 tweets are deleted from local cluster list.   
\item Full Cluster Evaluation: After local evaluation this bolt gets the previous clusters from Apache Cassandra database and again merges the local clusters and previous clusters held by database. For this step each global cluster came from database is evaluated for each local cluster one by one and the global cluster is updated locally (without affecting database) in case of similarity until all the similar local clusters are merged into this cluster. The merge decision is made by calculating the cosine similarity and the threshold of 0.5. This merge operation is also followed by the cluster word map re-evaluation step where words having smaller value than 0.01 as weight are deleted from map. After each similar local cluster is merged into the global cluster and weight updates, database entry of this global cluster is updated. In case of merge operation, this cluster also become an event candidate. For the updated global cluster, growth rate is calculated as explained above and if growth rate is bigger than the specified threshold which is 0.5, it is marked as event for the current round.
After merge operations, remaining local clusters are updated and added to database. Since new clusters contains very small number of tweets word map evaluation is made by the threshold value of 0.05. 
As final step global cluster elimination starts. The database entry of inactive clusters which are not updated for the last 3 blocks are deleted from database. 
\end{itemize} 
\item The Apache Storm topology of Twitter Event Detection by Clustering can be seen in Figure 4.
\newline

\end{itemize}

\section{EVALUATION}
For our experiments we have used the data collected for seven days from May 31, 2016 to April 7, 2016 and total number of tweets collected is nearly 12M. For each methodology described in the previous sections, we have compared the events and calculate similarity scores. Also we have compared the execution time of each methodology. 

\subsection{Process Times}
The list below summarizes the execution (process) times of each method. Since clustering methodology has many transactions with database to receive data and iterates over huge data, clustering methodology is the most time-consuming one. Keybased methodology which suspends streaming to prevent documents from mixing (results in 1680 suspensions) follows the clustering method by means of time consumption because this methodology uses some specified duration of sleeps for each document. Finally, the most efficient method is keybased event detection with direct grouping method due to the fact that this methodology has less iterations, works with smaller data blocks(words instead of tweets) and also does not use any suspensions.

\begin{table}[htb]
\centering
\caption{Execution Times}
\label{my-label}
\begin{tabular}{|l|l|l|l|}
\hline
Methods 									  & Exp1     & Exp2       & Exp3          \\ \hline
Keybased by shuffle grouping and suspensions  & 5.7h     & -          & -             \\ \hline
Keybased with direct grouping 				  & 3.1h     & 4.9h       & 5.02h         \\ \hline
Clustering method 							  & 8.9h     & 13.2h      & 13.06h        \\ \hline

\end{tabular}
\end{table}

\begin{table}[htb]
\centering
\caption{Number of Events}
\label{my-label}
\begin{tabular}{|l|l|l|l|}
\hline
Methods 									  & Exp1     & Exp2       & Exp3          \\ \hline
Keybased by shuffle grouping and suspensions  & 237      & -          & -             \\ \hline
Keybased with direct grouping 				  & 237      & 237        & 237           \\ \hline
Clustering method 							  & 297      & 263        & 290           \\ \hline

\end{tabular}
\end{table}

\subsection{Comparison of clustering and keybased methods}
Some statistics about the events found by keybased and clustering methods for Exp1:

\begin{itemize}
\item For CAN, 2 of the clusters include words detected by keybased methodology. Rate: 2/7 = 0.3
\item For USA, 24 of the clusters include words detected by keybased methodology. Rate: 24/290 = 0.083
\item For CAN, 6 of the words are included by clusters. Rate: 6/17 = 0.35.
\item For USA, 68 of the words are included by clusters. Rate: 68/220 = 0.309.
\end{itemize}


\addtolength{\textheight}{-12cm}   









\end{document}